\documentclass[twocolumn, superscriptaddress,  floatfix, amssymb,unsortedaddress, aps, pre, showpacs
]{revtex4}

\makeatletter
\def\@dotsep{4.5}
\makeatother

\def\mycaption#1{\caption{#1}}

\usepackage{graphics}
\usepackage[T1]{fontenc}
\usepackage[latin1]{inputenc}

\newcommand {\rsq}{\langle R^2 \rangle}

 \newcommand{\bu}{{\bf u}}

\begin{document}

\title{Viscoelasticity and primitive path analysis of entangled polymer liquids:
\\ From f-actin to polyethylene}

\author{Nariya Uchida}
\affiliation{Department of Physics, Tohoku University, Sendai 980-8578, Japan}

\author{Gary S. Grest}
\affiliation{Sandia National Laboratories, Albuquerque, NM 87185, USA}

\author{Ralf Everaers$^\ast$}
\affiliation{Universit\'e de Lyon, Laboratoire de Physique, \'Ecole Normale Sup\'erieure de Lyon, CNRS UMR 5672, 46 all\'ee d'Italie, 69364 Lyon Cedex 07, France}
\affiliation{Max-Planck-Institut f\"ur Physik komplexer Systeme, 
N\"othnitzer Str. 38, 01187 Dresden, Germany}

\date{\today}


\begin{abstract}   
We combine computer simulations and scaling arguments to develop a unified
view  of polymer entanglement based on the {\em primitive path analysis}
(PPA) of the microscopic topological state. Our results agree with experimentally measured plateau moduli  for three different polymer classes over a wide range
of reduced polymer densities: 
(i) semi-dilute theta solutions of synthetic polymers, 
(ii) the corresponding dense melts above the glass transition or crystallization temperature, and 
(iii) solutions of semi-flexible (bio)polymers such as f-actin or suspensions of rodlike
viruses. Together these systems  cover the entire range from loosely to tightly 
entangled polymers.
In particular, we argue that the primitive path analysis renormalizes a loosely to a tightly
entangled system and  provide a new explanation of the 
successful Lin-Noolandi packing 
conjecture for polymer melts.  
\end{abstract}


\maketitle


\section{Introduction}
The relation between the complex viscoelastic properties of
polymer liquids and their microscopic structure and dynamics is a key issue in
materials science and biophysics~\cite{doi86,mcleish02,boal02,Bent_sci_03,BauschKroy_natphys_06}.  
On a microscopic scale chains  can slide past each other, but their backbones 
cannot cross; the Brownian motion of these macromolecules
is hence subject to transient topological constraints~\cite{Edwards_procphyssoc_67},
an effect which is familiar from the manipulation of knotted strings. 
In slowing down the chain equilibration after a deformation, these constraints 
or entanglements
dominate the viscoelastic behavior of high molecular weight polymeric liquids.
Entanglement effects are universal, i.e. one observes the same
behavior for polymers with similar overall chain architecture (linear, ring, branched)
independently of details of the molecular structure.
Modern theories of polymer dynamics and rheology~\cite{doi86,mcleish02} 
describe the universal aspects of the viscoelastic behavior based on
the idea that molecular entanglements confine individual filaments to a
one-dimensional, diffusive dynamics (reptation) in tube-like regions in space.
Material specific parameters are determined through comparison to experiment.
Here we are concerned with the question, if these parameters and related
experimental observables can be inferred from the molecular structure of
polymeric liquids.

How strongly linear polymers entangle with each other depends on their
stiffness and on the contour length density of the polymer melt or
solution~\cite{graessley81}.   The microscopic structure is best discussed in
terms of the Kuhn length, $l_K$, and the number density of Kuhn segments,
$\rho_K$. The Kuhn length is  defined as the contour length $L$ where thermal
fluctuations start  to bend the chains and marks the crossover from rigid rod to
random coil behavior. In ``loosely'' \cite{morse98}
entangled systems with $\rho_K l_K^3< 1$ the
mean-free chain length between collisions is larger than the Kuhn length,
leading to random coil behavior between entanglement points. In contrast, for
$\rho_K l_K^3\gg1$  filaments are ``tightly''  \cite{morse98}
entangled and exhibit only small
bending fluctuations between entanglement points. 
As a consequence, the chains in tightly  entangled f-actin solutions
and in loosely entangled polyethylene melts behave differently on
the tube scale. The former are essentially stiff and
resist macroscopic shear due to an increase of their bending energy. The
latter are flexible and lose entropy when stretched locally.

\begin{figure}
\newpage
{\centering \resizebox*{\columnwidth}{!}{\rotatebox{-0}
{\includegraphics{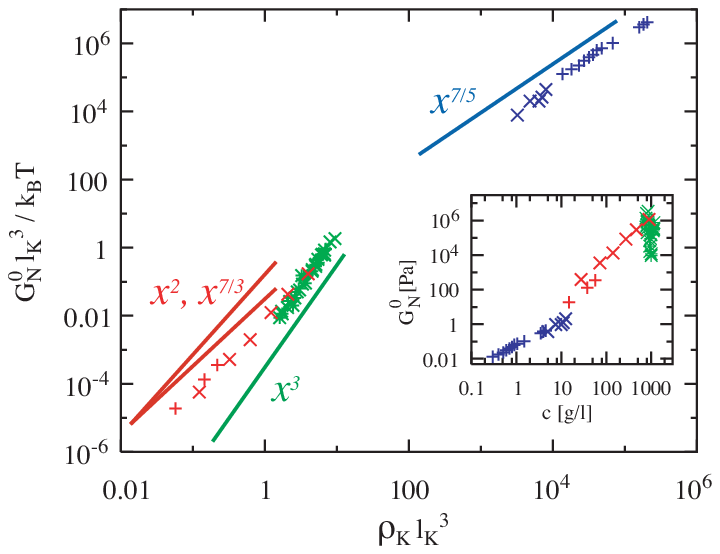}}} }
\label{fig:GExpScaling}
\mycaption{
  Dimensionless plateau moduli $G_N^0\,l_K^3/k_BT$ as a function of the
  dimensionless number density of Kuhn segments $\rho_K l_K^3$:
  Experimental data vs. scaling predictions.
  Symbols indicate experimental data 
  for (i) tightly entangled f-actin (blue $+$)~\cite{HinnerSackmann_prl_98} and 
  fd-phage (blue $\times$)~\cite{SchmidtSackmann_pre_00} solutions, (ii)
  various loosely entangled polydiene, polyolefine, and 
  polyacrylate melts (green $\times$)~\cite{fetters94,Sukumaran_PPA_05}, and loosely 
  entangled theta-solutions  of polystyrene (red $+$) \cite{inoue02} and polybutadiene(red $\times$) 
  \cite{colby91}. 
  Lines represent theoretically derived power laws 
  $G_N^0\propto \left(\rho_K l_K^3\right)^\alpha$ for tightly~\protect\cite{Semenov_far_87} and loosely~\protect\cite{colby92} entangled systems.
  }
\end{figure}

The differences between these
two situations become apparent, when one considers the relation between the
microscopic solution structure and the height of the characteristic rubber-like plateau 
in the shear relaxation modulus, $G_N^0$.
In Fig.~\ref{fig:GExpScaling} we show
a Graessley-Edwards plot~\cite{graessley81} of
experimentally measured~\cite{HinnerSackmann_prl_98,SchmidtSackmann_pre_00,fetters94,Sukumaran_PPA_05, inoue02, colby91}
dimensionless plateau moduli $G_N^0 l_K^3/k_BT$ as a function of the
dimensionless number density of Kuhn segments $\rho_K l_K^3$. 
For comparison, the unscaled plateau moduli are shown 
in the inset as a function of polymer concentration.
The data we have compiled 
represent the behavior of several
prototypical classes of entangled polymers:  (i) tightly entangled solutions of
semi-flexible biopolymers such as  f-actin and suspension of fd-phages with
$\rho_K l_K^3\gg 10$, $l_K\approx 10^{-6} m$ and $G_N^0\approx 10^{-1}$Pa,  and
(ii) loosely entangled melts of commercially important, synthetic polymers with
$\rho_K l_K^3<10$, $l_K\approx 10^{-9} m$ and plateau moduli of the order of
$G_N^0\approx 10^6$Pa. Typical synthetic polymers are sufficiently flexible to
be in the isotropic phase in the melt state with a volume fraction $\Phi =
\rho_K\, l_K\, d^2={\cal O}(1)$.  In contrast, tightly entangled chains have to
have a sufficiently small diameter $d$ to fulfill the Onsager criterion $\Phi
(l_K/d)<5$ for the isotropic-to-nematic transition. We note that the crossover
between the two regimes in Fig.~\ref{fig:GExpScaling} is located close to the threshold
for the isotropic-nematic transition in dense polymer melts~\footnote{   The
location of the isotropic-nematic transition is given by the Onsager criterion
$\rho_K\, l_K^2\,d\approx5$. Dense melts with polymer volume fractions $\Phi=
\rho_K\, l_K\, d^2 ={\cal O}(1)$ are isotropic for sufficiently flexible chains
with aspect ratios $l_K/d<5$ or  $\rho_K l_K^3<25$. In contrast, isotropic,
tightly entangled solutions  with $ \rho_K l_K^3\gg1$ can only be formed by
semi-flexible chains with large aspect ratios $l_K/d\gg(\rho_K l_K^3)^{1/2}$.}.
Furthermore, Fig.~\ref{fig:GExpScaling} contains data for  (iii) loosely entangled
semi-dilute solutions of synthetic polymers in so called theta-solvents. While
these systems are supposed to preserve the chain conformational  statistics from
the undiluted melt, the observed reduced plateau moduli are larger than those
for dense systems and exhibit a  qualitatively different density dependence.

For tightly entangled systems, the relation between the entanglement density,
$\rho_e$, and $\rho_K l_K^3$ was determined by Semenov~\cite{Semenov_far_87}
from a geometrical argument: the area swept out via transverse fluctuations by a
filament between two entanglement points is on average traversed by one other
filament serving as an obstacle. In loosely entangled systems it is impossible
to determine $\rho_e$ by scaling arguments
alone~\cite{deGennes74,Brochard77,graessley81,lin87,kavassalis87,Semenov_far_87,colby90,colby92,milner_mm_05}. 
A promising tool to solve this problem from first principles is the primitive
path
analysis (PPA)~\cite{PPA,Sukumaran_PPA_05,Kroeger_cpc_05,ShanbhagLarson_prl_05,ZhouLarson_mm_05,Tzoumanekas_mm_06}.
Primitive paths were originally introduced in a thought experiment to determine
the tubes confining individual polymers in an entangled polymer melt or
network.  The idea is to identify the random walk-like tube axis with the
shortest (``primitive'') paths between the end points of the original chains into
which the chain contours can be contracted without crossing each
other~\cite{edwards77,RubinsteinHelfand_jcp_85}.  
As we have shown~\cite{PPA,Sukumaran_PPA_05}, the numerical implementation of
this idea allows to make quantitative predictions of melt
viscoelastic properties on the basis of a topological analysis. Here  we
(i) argue that the PPA can be understood as a way of renormalizing a loosely
to a tightly entangled system,
(ii) derive a modified relation between the 
primitive path mesh characterizing an entangled polymer liquid and
its macroscopic properties (Sec.~\ref{sec:theory}),
(iii) apply the PPA to a much wider range of loosely and tightly entangled model
polymer structures (Sec.~\ref{sec:PPA}), 
(iv) validate our results by a comparison to the experimental data displayed
in Fig.~\ref{fig:GExpScaling}, and
(v) discuss the relation between the characteristic length scales of the primitive
path mesh and the packing length (Sec.~\ref{sec:results}).
We conclude with a brief summary in Sec.~\ref{sec:summary}.

\section{Theory}\label{sec:theory}

The result of the PPA is a mesh of mutually
entangled, piecewise straight primitive paths
(see, e.g. Fig.~3 in Ref.~\cite{PPA}). The structure can be
characterized by the contour length, $L_{pp}$, the mesh size,
$\xi_{pp}=1/\sqrt{\rho_{chain} L_{pp}}$, and the Kuhn length, $a_{pp}$, of the
primitive paths~\cite{PPA,Sukumaran_PPA_05}.  Furthermore, it is useful to
introduce the average contour length, $l_e$, of the primitive paths between
entanglement points as well as the corresponding chain contour length, $L_e$. On a
scaling level, the latter is defined implicitly by the relation

\begin{equation}\label{eq:le_def}
l_e^2 = \rsq(L_e,l_K)
\end{equation}
where $\rsq(L,l_K)$ denotes the mean-square spatial distance of two
points separated by a length $L$ along the contour. The chain and
primitive path statistics agree asymptotically~\cite{doi86}, 
\begin{math}
\lim_{L\rightarrow\infty}\rsq = l_K L \equiv a_{pp} L_{pp} = 
\lim_{L_{pp}\rightarrow\infty}\rsq_{pp}
\end{math}
so that the Kuhn length 

\begin{equation}\label{eq:app}
\frac{a_{pp}}{l_K}= \frac1{(l_e/L_e)}
\end{equation}
of the primitive paths increases by the inverse of the shrinking factor
of the contour length.  Similarly the primitive path mesh size is given by

\begin{equation}\label{eq:xipp}
\frac{\xi_{pp}}{l_K}=\frac1{(\rho_K l_K^3)^{1/2} (l_e/L_e)^{1/2}}.
\end{equation}

In the following, we assume (i) that all information necessary to calculate
the plateau modulus of a sample can be deduced from the primitive
path mesh characterizing its microscopic topological state
and (ii) that the {\em primitive path} structure -- 
viscoelastic property relation is system independent, i.e. after carrying
out the PPA it is no longer necessary to distinguish between different
polymer classes. Moreover, we note that by construction primitive
path meshes resemble tightly entangled solutions of semi-flexible
chains and that the latter are invariant under the PPA. 

Regarding the primitive path analysis as a means to renormalize a 
loosely to a tightly entangled system, allows us to
adapt two results from the theory of tightly entangled solutions of
semi-flexible chains \cite{Semenov_far_87} to the present situation:
\begin{eqnarray}
\xi_{pp}^2 &=& c_\xi\ l_e \times \left(l_e^{3}/a_{pp}\right)^{1/2}
\label{eq:le_cond}\\
\frac{G_N^0\, l_K^3}{k_BT} &=& 
c_G \left(\frac{l_K}{\xi_{pp}}\right)^2\frac{l_K}{l_e}
\nonumber\\
&=&c_G c_\xi^{2/5} (\rho_K l_K^3)^{7/5} (l_e/L_e)^{8/5}
\label{eq:Gpp}
\end{eqnarray}
Eq.~(\ref{eq:le_cond}) expresses the idea that the area swept out via
transverse fluctuations by a primitive path between two entanglement points is
on average traversed by one other primitive path serving as an obstacle.
Eq.~(\ref{eq:Gpp}) states that the plateau modulus is proportional to
$k_BT$ times the density of entanglement points.

For given chain statistics it is possible to determine $l_e$
self-consistently from Eqs.~(\ref{eq:le_def}-\ref{eq:le_cond}). 
For tightly entangled systems,  our approach reduces by construction
to the standard results for semi-dilute solutions of semi-flexible
chains~\cite{Semenov_far_87,morse_01}:
$a_{pp}/l_K=1$, $\xi_{pp}/l_K=1/(\rho_K l_K^3)^{1/2}$,
$l_e/l_K\propto 1/(\rho_K l_K^3)^{2/5}$
and $G_N^0\, l_K^3/(k_BT)\propto(\rho_K l_K^3)^{7/5}$.
In the opposite limit, we find
$a_{pp}/l_K\propto \xi_{pp}/l_K\propto1/(\rho_K l_K^3)$
and $G_N^0\, l_K^3/(k_BT)\propto(\rho_K l_K^3)^3$
i.e., we have presented a {\em derivation} of the Lin-Noolandi
conjecture~\cite{lin87,kavassalis87}.
In the general case of solutions where the individual polymers 
exhibit worm-like chain statistics on all length scales, 
a reasonable approximation (indicated by solid lines in all
figures) is given by 

\begin{eqnarray}
\label{eq:le_approx}
\frac{l_e}{L_e} &\approx&
\left(1+ \frac{L_e}{l_K}\right)^{-1/2}\\
\frac{L_e}{l_K} &\approx&
\left(\frac1{c_\xi(\rho_K l_K^3)}\right)^{2/5} +
         \left(\frac1{c_\xi(\rho_K l_K^3)}\right)^{2}.
\label{eq:Le_approx} 
\end{eqnarray} 
By inserting Eqs.~(\ref{eq:le_approx}) and (\ref{eq:Le_approx}) into Eqs.~(\ref{eq:app}),
(\ref{eq:xipp}), and (\ref{eq:Gpp}) we thus arrive at a prediction 
for the dependence of the
properties of the primitive path mesh on the dimensionless Kuhn length
density $\rho_K l_K^3$ of the original polymer melt or solution.

\begin{figure}[t]
{\centering \resizebox*{1\columnwidth}{!}{\rotatebox{-0}
{\includegraphics{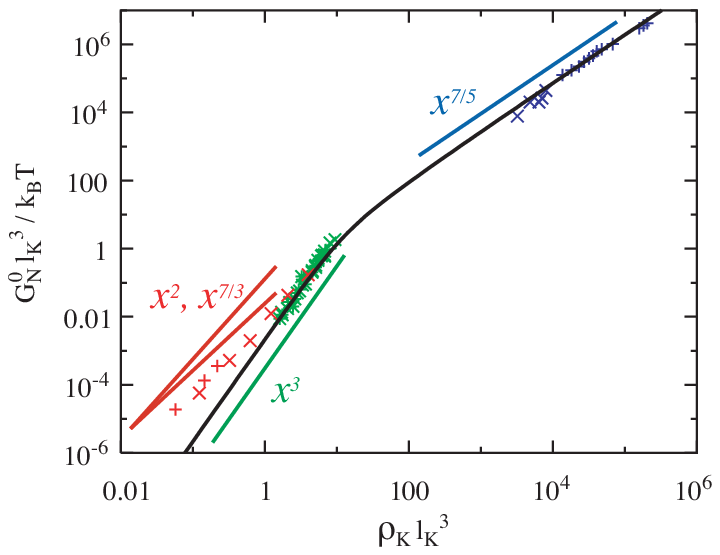}}} }
\label{fig:GExpTheo}
\mycaption{
  Dimensionless plateau moduli $G_N^0\,l_K^3/k_BT$ as a function of the
  dimensionless number density of Kuhn segments $\rho_K l_K^3$:
  Theory vs. experimental and scaling results as in Fig.~\protect\ref{fig:GExpScaling}.
  The black solid line indicates the combination of our Eqs.~(\protect\ref{eq:Gpp}) 
  and (\protect\ref{eq:le_approx}) with $c_\xi=0.06$ and $c_G=0.6$.
  }
\end{figure}

Choosing $c_\xi=0.06$ and $c_G=0.6$ we find excellent agreement between our
theory and the experimental data for tightly entangled solutions and loosely
entangled dense melts (Fig.~\ref{fig:GExpTheo}). 
However, the Lin-Noolandi conjecture (and hence also
our ansatz) fail for loosely entangled theta-solutions which seem to be better
described  by the theory of Colby and Rubinstein~\cite{colby90,colby92}.  An 
important oversimplification was
recently pointed out by Milner~\cite{milner_mm_05}:  by accounting only for the
{\em mean} segment density, one  implicitly treats the solvent as providing a
``coating'' which increases the effective chain  diameter to $d/\sqrt{\Phi}$. As
consequence, one neglects the increased probability of entanglement formation
between closely approaching chain sections.

\section{Primitive Path Analysis}\label{sec:PPA}

Are these effects preserved in an explicit primitive path analysis?
To answer this question we have generated and analyzed model polymer
liquids corresponding to the experimental data: (i) tightly entangled solutions
of zero-diameter wormlike chains (WLCs) with $10<\rho_K l_K^3<10^5$, 
(ii) dense melts of flexible
bead-spring chains with $\Phi = O(1)$ and $1<\rho_K l_K^3<40$, and (iii) model
theta-solutions generated by eliminating a fraction $0.1<\Phi<0.925$ of the
chains from equilibrated~\cite{Auhl_jcp_03} conformations of dense bead-spring
chain melts.  

\subsection{Bead-spring polymer solutions and melts}

For the loosely entangled regime we used the model and procedure described in
Refs.~\cite{PPA,Sukumaran_PPA_05}. Monomers are modeled as spheres of diameter
$\sigma$ interacting through a purely repulsive 6-12 Lennard-Jones (LJ) potential,
which is short ranged and purely repulsive.  The polymers are formed by
connecting beads via non-linear FENE springs. The average bond length is $b=0.97
\sigma$. The parameter choice ensures that two chains cannot cross each other in
dynamic simulations.  Monodisperse polymer melts of $ M=80-500$ chains of length
$50 \le N \le 700$ at a bead density of $\rho =0.85\sigma^{-3}$ are studied. By
introducing a small intrinsic bond bending potential, $l_k$ is varied between
$1.82 \sigma$ and $3.34 \sigma$, for details see Ref.\cite{Auhl_jcp_03}. For dense
systems, we extended the data from Ref.~\cite{PPA} by one additional data point
for larger intrinsic bending stiffness close to the isotropic nematic
transition. Furthermore, we equilibrated two Kremer-Grest theta solutions 
$ M=200$ chains of length
$N=700$ at a bead density of $\rho =0.25\sigma^{-3}$ and 
$\rho =0.4\sigma^{-3}$. In this case, the LJ cutoff is set to $r_c=2.5\sigma$
and simulations are carried out at $k_BT=3.0\epsilon_{LJ}$.
To obtain a large number of
model theta-solutions whose intra- and inter-chain
correlations are {\em identical} to those of the dense systems, we 
randomly eliminate a fraction $0.1<\Phi<0.925$ of the chains from 
equilibrated \cite{Auhl_jcp_03} conformations of dense
bead-spring chain melts with $N=7000$. The PPA is implemented into a standard
Molecular Dynamics code~\cite{PPA,Sukumaran_PPA_05}: Chain ends are fixed in space,  intra-chain excluded
volume as well as bending interactions are disabled, and chain contraction is
induced by  cooling the system toward $T=0$.

\subsection{Entangled solutions of zero-diameter WLCs}

Data covering the crossover to the tightly entangled regime were mainly obtained using
Monte Carlo techniques. We have generated and analyzed semi-dilute solutions of
infinitely thin worm-like chains with $1 \le \rho_K l_K^3 \le 10^5$.  Bending of
the chain was penalized by the Hamiltonian \begin{math} H/k_BT
=- (l_p/b) \sum_{i=1}^{L-1} \bu_i \cdot \bu_{i+1}, \end{math}  where $b$
denotes the segment length, $l_p=\kappa/k_BT$ is the persistence length of the
continuum worm-like chain with the bending modulus $\kappa$, and $\bu_i$ the
unit vector along the axis of the $i$-th cylinder.  Chains of this type have a
Kuhn length of  \begin{math}\label{eq:lK}  l_K^{(0)}/b = 2/(1 + b/l_P
-\coth l_P/b) -1  \end{math}  and can be efficiently generated by simple
sampling. In the absence of inter-chain correlations for zero chain diameter,
fully equilibrated semi-dilute solutions are obtained by placing $N$ randomly
generated chains into a cubic box of size $L_{box}^3$ with periodic boundary
conditions. As a general rule, the chains were chosen to be long enough to be
multiply entangled and the linear dimension of the simulation box exceeds the
typical chain extensions.  For the PPA we used a Brownian Dynamics/force-biased
Monte-Carlo simulation code for entangled worm-like chains, which rejects moves
leading to chain crossing. Starting from the solution conformations, the
positions of the chain ends are fixed. Chain contraction is induced by setting
chain stiffness as well as the equilibrium extension of the segments to zero and
reducing the temperature.

\subsection{Estimating plateau moduli from the PPA}
\label{sec:GDoi}

Following again Refs.~\cite{PPA,Sukumaran_PPA_05} we measure mean-square
internal distances $\langle r^2_{pp}(|i-j| \rangle$ as a function of chemical
distance $|i-j|$ to determine the Kuhn length, $a_{pp}$, and the contour 
length, $L_{pp}$ of the primitive paths. 

For flexible polymers the standard relation between the plateau modulus and the
length scales characterizing the primitive path mesh is~\cite{doi86}

\begin{equation}
\frac{G_N^0\, l_K^3}{k_BT} \propto 
(\rho_K l_K^3) \left(\frac{l_k}{a_{pp}}\right)^{2}
\label{eq:GppDoi}
\end{equation}
Here we have instead used Eq.~(\ref{eq:Gpp}) which
can be written as

\begin{eqnarray}
\frac{G_N^0\, l_K^3}{k_BT} &=& 
c_G c_\xi^{2/5} (\rho_K l_K^3)^{7/5} \left(\frac{l_k}{a_{pp}}\right)^{8/5}
\label{eq:Gpp1}
\end{eqnarray}
Eqs.~(\ref{eq:GppDoi}) and (\ref{eq:Gpp1}) are equivalent, if
$a_{pp}/l_K\propto \xi_{pp}/l_K\propto1/(\rho_K l_K^3)$, i.e.
for the dense, loosely entangled melts where Eq.~(\ref{eq:GppDoi}) was used in
Ref.~{PPA}. However, Eq.~(\ref{eq:GppDoi}) clearly fails in the 
tightly entangled regime and we also found less good agreement with experimental data
for estimates of plateau moduli for theta-solutions (data not shown).

\section{Results and Discussion}\label{sec:results}

\begin{figure}[t]
{\centering \resizebox*{\columnwidth}{!}{\rotatebox{-0}
{\includegraphics{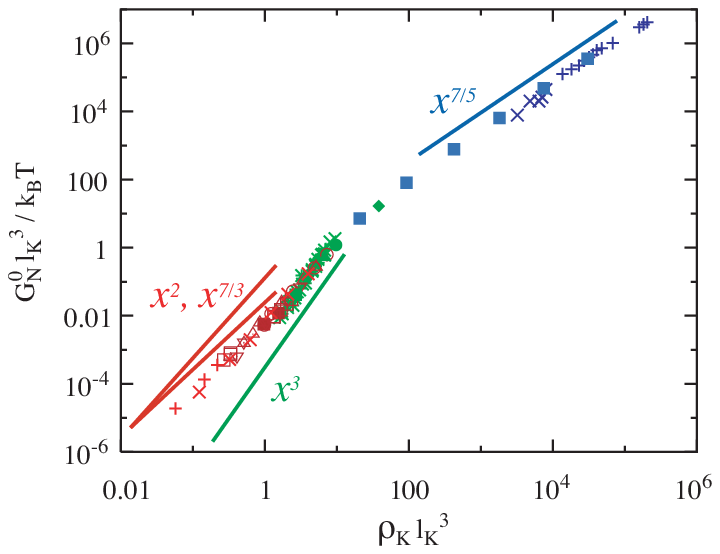}}} }
\label{fig:GPPA}
\mycaption{
  Dimensionless plateau moduli $G_N^0\,l_K^3/k_BT$ as a function of the
  dimensionless number density of Kuhn segments $\rho_K l_K^3$:
  PPA vs. experimental and scaling results as in Fig.~\protect\ref{fig:GExpScaling}.
  Additional symbols indicate PPA results 
  for 
  (i) tightly entangled, zero-diameter worm like chains (blue $\square$), (ii)
  Kremer-Grest melts and theta-solutions (red and green filled symbols) 
  for bead-spring polymers with intrinsic stiffness 
  $\square < \bigtriangleup < \bigtriangledown < \bigcirc < \Diamond$.
  Furthermore, we have included PPA results for model 
  theta-solutions (open red symbols) created by
  eliminating chains from the corresponding melts.  
  }
\end{figure}

In Fig.~\ref{fig:GPPA} we show our PPA results for the plateau moduli in 
comparison to the experimental data. The PPA identifies the location of the
crossovers from the theta-solution to the dense melts and from the loosely to
the tightly entangled regime with no adjustable parameters.  In particular, we
obtain {\em quantitative} agreement with the experimentally measured moduli by
adjusting a  {\em single} parameter (equal to $c_G c_\xi^{2/5}=0.2$)
for the strength of the elastic response. Before we take a closer look at
the results for loosely entangled theta solutions, we discuss the microscopic
length scales characterizing the primitive path mesh.

\begin{figure}[t]
{\centering \resizebox*{\columnwidth}{!}{\rotatebox{0}
{\includegraphics{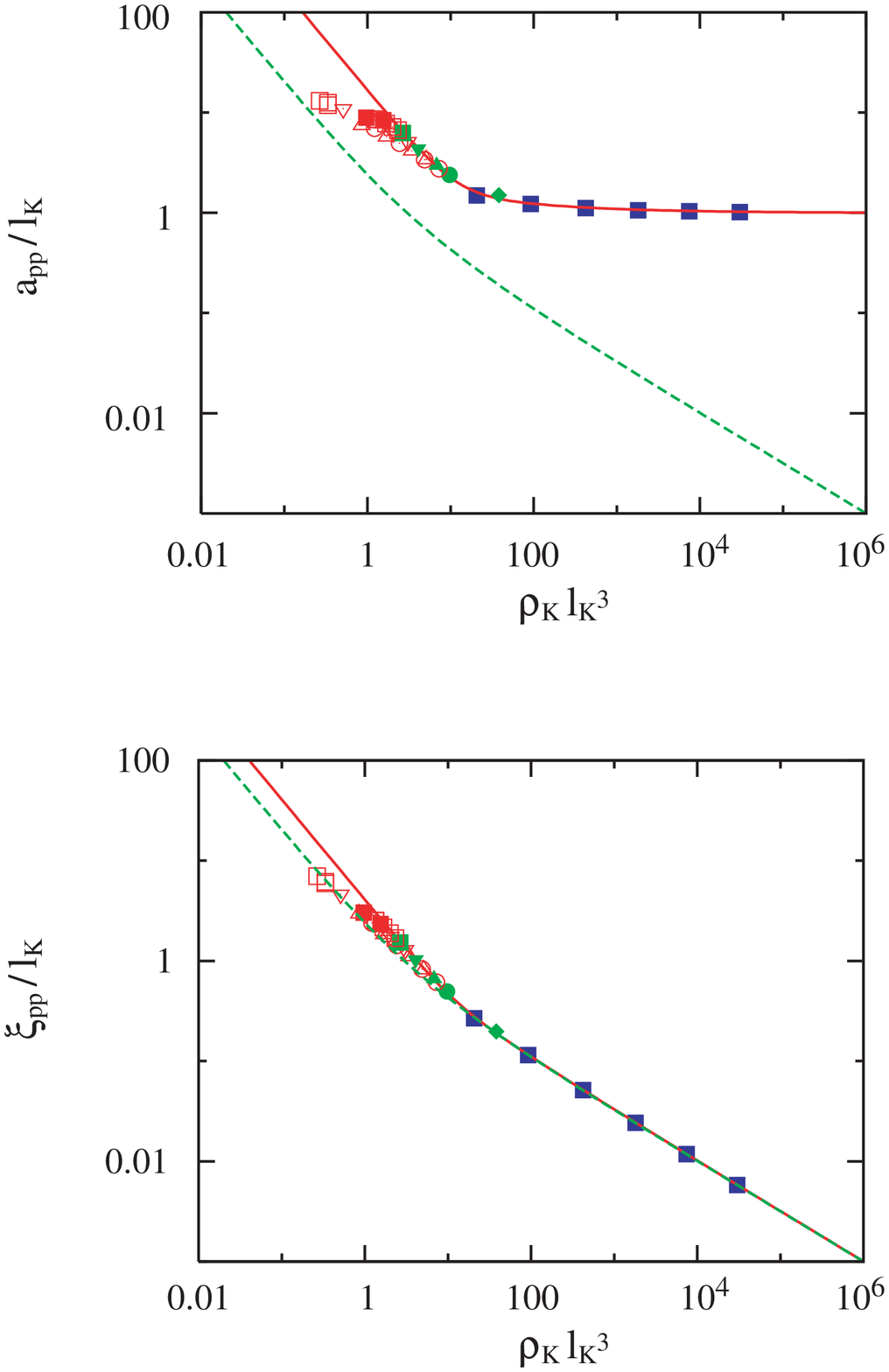}}} }
\label{fig:app_xipp}
\mycaption{
  Kuhn length $a_{pp}/l_K$ and mesh size $\xi_{pp}/l_K$  for
  the primitive path mesh as a function of the dimensionless number
  density of Kuhn segments $\rho_K l_K^3$. 
  The solid lines indicate the combination of Eq.~(\protect\ref{eq:le_approx})
  with Eqs.~(\protect\ref{eq:app}) and (\protect\ref{eq:xipp})
  respectively with $c_\xi=0.06$ and $c_G=0.6$. The dashed line
  represents the packing length Eq.~(\protect\ref{eq:p_approx}).
  Symbols indicate
  PPA results for zero-diameter wormlike chains (blue), dense bead-spring
  melts (green) and model theta-solutions (red).  }
\end{figure}

In Fig.~\ref{fig:app_xipp} we show the Kuhn lengths $a_{pp}$ and mesh sizes
$\xi_{pp}$ which we extracted for the various model systems. As for the
predicted moduli, the results for tightly entangled WLC solutions and for dense
melts of flexible bead-spring chains are in excellent agreement with our
theoretical results. Note that there are no adjustable parameters in
addition to the values of $c_\xi=0.06$ and $c_G=0.6$ derived above from the
experimentally measured  plateau moduli. 

Interestingly, the PP mesh size $\xi_{pp}$ is essentially given by  the packing
length, $p$. This length scale was introduced~\cite{witten89b} as the
characteristic spatial distance where different polymers start to
interpenetrate, i.e. the length scale where the intra- and interchain monomer
pair correlation functions coincide: $g_{intra}(p) \equiv g_{inter}(p)$. In the
systems we investigate interchain correlations are small, so that we use
$g_{inter}(p) \equiv 1$ or 
\begin{math}\label{eq:pdef}  
\frac{\partial
L(p,l_K)}{\partial p} \equiv p^2 \rho_K l_K  
\end{math} 
to calculate the packing length. Using  

\begin{eqnarray}
\rsq(L,l_K)  &\approx&  L^2/(1+L/l_K)
\label{eq:rsq_appr}\\
L(r,l_K)  &\approx&  r(1+r/l_K) 
\end{eqnarray}
the result is

\begin{equation}\label{eq:p_approx}
\frac p{l_K} \approx \frac{1+\sqrt{1+(\rho_Kl_K^3)}}{(\rho_Kl_K^3)}
\end{equation}
The limiting cases are again simple to understand.  For $p\gg l_K$, the chain
contour length $L$ contained in a volume $p^3$ is much longer than the Kuhn
length $l_K$. The chains follow a random walk statistics with $\rsq=l_K L$. In
this case, the packing length is given by $p=(\rho_{chain}\rsq)^{-1}$ 
~\cite{witten89b} and with
$p/l_K\propto a_{pp}/l_K\propto \xi_{pp}/l_K\propto1/(\rho_K l_K^3)$ the
packing length turns out to be the only relevant length scale in loosely
entangled systems~\cite{fetters94,PPA}.  In the opposite limit, when $p\ll l_K$
the single chain statistics corresponds to a rigid rod with $\rsq=L^2$. In
this case, $p=(\rho_{chain}\rsq^{1/2})^{-1/2}=1/(\rho_K l_K)^{1/2}$, i.e. $p$
is identical to the mesh size of the solution as well as the primitive path
mesh. In contrast to the previous case, the Kuhn length $a_{pp}=l_K$ of both
chains and primitive paths remains as a second, independent length scale.

As a last point, we come back to the case of loosely
entangled (model) theta-solutions. Both, experimental and PPA
results for plateau moduli show clear deviations from 
our theory (respectively the packing conjecture)
$G_N^0\, l_K^3/(k_BT)\propto(\rho_K l_K^3)^3$
and the corresponding extrapolation of the
behavior of loosely entangled dense melts (Fig.~\ref{fig:GPPA}).
Qualitatively, sufficiently dilute systems behave in accordance with
a power law $G_N^0\, l_K^3/(k_BT)\propto(\rho_K l_K^3)^\alpha$
where the exponent $\alpha$ is in the range of the predictions
~\cite{colby92} 
of the binary contact model ($\alpha=2$) \cite{deGennes74,Brochard77} and the 
Colby-Rubinstein model ($\alpha=7/3$)~\cite{colby90,milner_mm_05}. 
However, for the relatively dense model theta solutions
which we analyzed using the PPA, we see no evidence 
for immediate departures from the packing line of the
form $G_N^0(\Phi)=G_N^0(\Phi=1)\Phi^\alpha$, i.e. for a family
of curves with identical slope but different prefactor 
essentially determined by the plateau moduli of the undiluted melt.
Rather, data obtained for model theta solutions map fairly
well onto each other and continue to map onto
the melt results as long as $\rho_K l_K^3>1$. Deviations only occur
for lower concentrations.  

The comparison can be extended to the   length scales characterizing the
primitive path meshes. Inserting the predictions for $N_e\equiv L_e/l_K$
of the binary contact and the Colby-Rubinstein model into Eqs.~(\ref{eq:app})
and (\ref{eq:xipp}) (which follow directly from the definitions and hold 
independently of the applicability of the Semenov theory to the primitive
path mesh) yields 
$a_{pp}/l_K \propto \left( \rho_K l_K^3 \right)^{-1/2}$ ,
$\xi_{pp}/l_K \propto \left( \rho_K l_K^3 \right)^{-3/4}$ and
$a_{pp}/l_K \propto \left( \rho_K l_K^3 \right)^{-2/3}$ ,
$\xi_{pp}/l_K \propto \left( \rho_K l_K^3 \right)^{-5/6}$ respectively.
In particular, one obtains for the ratio
$a_{pp}/\xi_{pp} \propto \left( \rho_K l_K^3 \right)^{1/4}$ and
$a_{pp}/\xi_{pp} \propto \left( \rho_K l_K^3 \right)^{1/6}$ respectively.
Thus, in the limit in question where $\left( \rho_K l_K^3 \right)$ can become
arbitrarily small, these models predict that the area $a_{pp}^2$ mapped
out by two subsequent Kuhn segments of a primitive path is on average
traversed by {\em less than one} other primitive path which could
serve as an obstacle. The results of the actual PPA shown in
Fig.~\ref{fig:app_xipp} allow no definite conclusion. Naive extrapolation
would indeed lead to $a_{pp}<\xi_{pp}$ in the limit 
$\left( \rho_K l_K^3 \right)\rightarrow0$. However, we note that we have
never observed this inversion. Interestingly, the deviations
from our theory occur once $p$ exceeds $l_K$ and 
the results of the explicit PPA for theta-solutions  are
compatible with a crossover to a pure packing scenario with
$p/l_K = a_{pp}/l_K = \xi_{pp}/l_K\propto1/(\rho_K l_K^3)$ where
the PP directions before and after an entanglement point become
completely uncorrelated.

\section{Summary and conclusion}\label{sec:summary}

There is a wide spectrum of entangled polymer liquids
whose single chain structure is characterized by a single microscopic length scale, the 
Kuhn or persistence length. Our starting point was a compilation of experimentally
measured plateau moduli for 
semi-dilute theta solutions of synthetic polymers, the corresponding dense melts above the glass transition or crystallization temperature, and solutions of semi-flexible (bio)polymers such as f-actin or suspensions of rodlike
viruses. Together these systems  cover the entire range from loosely to 
tightly entangled polymers.
In this article, we have demonstrated
excellent agreement between the experimentally measured plateau moduli 
and our results derived from a 
primitive path analysis~\cite{PPA} of corresponding model polymer liquids.
This is strong evidence supporting our working hypothesis that the PPA may be
regarded as a tool to renormalize a loosely to a tightly entangled system:
while the relation between the microscopic structure and the viscoelastic
properties is different for the three classes of entangled polymer liquids
included in the present study, these differences vanish in the course of
the PPA. In particular, experimental properties can be calculated 
by applying relations for tightly entangled systems to the primitive path mesh
{\em independently} of the character of the original system 
(note that Eq.~(\ref{eq:Gpp}) for the plateau modulus reduces 
to the standard expression~\cite{doi86} Eq.~(\ref{eq:GppDoi}) in the case of loosely entangled melts; however, the latter {\em fails} in the tightly entangled regime).
This ansatz provided a new
explanation of the Lin-Noolandi packing conjecture~\cite{lin87,kavassalis87} and
allowed us to derive simple analytical expressions for macroscopic
(Fig.~\ref{fig:GExpScaling}) and microscopic (Fig.~\ref{fig:app_xipp}) entanglement properties which are in excellent agreement
with experimental and simulation data over a wide range of reduced polymer
densities. Future work has to show whether  the
Colby and Rubinstein scaling~\cite{colby90} in theta-solutions
is a cross-over effect
or valid asymptotically and whether  it is possible to account for
density fluctuations by combining the arguments and methods presented in this
paper  with the scaling analysis of loosely entangled solutions presented in
Ref.~\cite{milner_mm_05}. Judging from the available experimental
and PPA results, this might be
necessary to ``disentangle'' the influence of the various crossovers in the
experimentally relevant parameter range.
\\
\\
\acknowledgments
We acknowledge helpful discussions 
with K. Kremer, S. Sukumaran, and C. Svaneborg.  NU acknowledges financial support by Grand-in-Aid for
Scientific Research from Japan's Ministry of Education, Culture, Sports,
Science and Technology, and the hospitality of the Max-Planck-Institute for
Polymer Research in Mainz where we started to develop the simulation
code. Sandia is a multiprogram laboratory operated by Sandia Corporation, a
Lockheed Martin Company, for the United States department of Energy's National
Nuclear Security Administration under contract de-AC04-94AL85000.
RE is supported by the chair of excellence program of the French Agence 
Nationale de Recherche.

\bibliography{Entanglement,poly,ref13,solbib,degennes74}

\end{document}